\documentclass[12pt]{iopart}
\usepackage{graphics}
\usepackage{url}

\begin{document}
\title[Symmetry Energy Effects on the Radii and Tidal Polarizabilities of Neutron Stars]
{Delineating Effects of Nuclear Symmetry Energy on the Radii and Tidal Polarizabilities of Neutron Stars}

\author{Nai-Bo Zhang$^{1,2}$ and Bao-An Li$^{2*}$}

\address{$^{1}$Shandong Provincial Key Laboratory of Optical Astronomy and Solar-Terrestrial Environment, School of Space Science and Physics,
Institute of Space Sciences, Shandong University, Weihai, 264209, China}
\address{$^{2}$Department of Physics and Astronomy, Texas A$\&$M University-Commerce, Commerce, TX 75429, USA}
\ead{$^*$Corresponding author:Bao-An.Li@Tamuc.edu}
\vspace{10pt}

\begin{abstract}
What can we learn about the density dependence of nuclear symmetry energy $E_{\rm{sym}}(\rho )$ from precise measurements of the radius ($R_{\rm{1.4}}$) and/or tidal polarizability ($\Lambda_{1.4}$) of canonical neutron stars (NSs) with a mass of 1.4 M$_\odot$? With the $E_{\rm{sym}}(\rho )$ parameterized using three parameters $L$, $K_{\rm{sym}}$, and $J_{\rm{sym}}$ which have the asymptotic meaning of being respectively the slope, curvature, and skewness of symmetry energy at saturation density, we found that, while both the $R_{\rm{1.4}}$ and $\Lambda_{1.4}$ depend strongly on the slope $L$, the $K_{\rm{sym}}$ and $J_{\rm{sym}}$ parameters characterizing the high-density behavior of $E_{\rm{sym}}(\rho )$ also play appreciable roles. Thus, there is not a simple relation between the $\Lambda_{\rm{1.4}}$/$R_{\rm{1.4}}$ and $L$ alone. Precise measurements of just the $\Lambda_{\rm{1.4}}$ and $R_{\rm{1.4}}$ can not completely determine the $E_{\rm{sym}}(\rho )$ but limit combinations of its parameters. In particular, stringent constraints approximately independent of the $J_{\rm{sym}}$ on the $L$-$K_{\rm{sym}}$ correlations can be obtained. However, infinite combinations of the larger (smaller) $L$ and smaller (larger) $K_{\rm{sym}}$ can lead to the same $\Lambda_{\rm{1.4}}$ and $R_{\rm{1.4}}$. Additional observables including those from terrestrial nuclear experiments are thus necessary to break this degeneracy in order to completely determine the density dependence of nuclear symmetry energy $E_{\rm{sym}}(\rho )$.

\end{abstract}
\submitto{\jpg}
\maketitle

\section{Introduction}\label{intro}
The first detection of gravitational waves (GW) from a binary neutron star (NS) merger event  GW170817 \cite{LIGO1,LIGO2017,LIGO2018} has created an enormous new wave of detailed studies on the Equation of State (EOS) of dense neutron-rich matter in both astrophysics and nuclear physics communities. In particular, new constraints on the radii of canonical NSs and the associated EOS of dense neutron-rich matter have been inferred from the reported tidal polarizability of NSs in GW170817 in a welcomed flood of interesting papers . For example, the LIGO+Virgo Collaborations recently inferred a radius of $R_{\rm{1.4}}$= 10.5-13.3 km using the dimensionless tidal polarizability $\Lambda_{\rm{1.4}}=70-580$ from their refined analyses of GW170817 \cite{LIGO2018}. Other analyses using sometimes different techniques and EOSs based on various nuclear many-body theories and interactions
have found a rather consistent upper limit of $R_{\rm{1.4}}\leq 13.7$ km \cite{Ang-mass,Ann,Fattoyev17,Most,Plamen3,Rai,Tew18,Malik,Holt18} using the maximum value of $\Lambda_{\rm{1.4}}\le 800$ first reported in \cite{LIGO1}. In turn, either the tidal polarizability $\Lambda_{\rm{1.4}}$ itself and/or the radius $R_{\rm{1.4}}$ has been used to constrain the underlying EOS used in the model analyses.

In the ongoing efforts of constraining quantitatively details of the EOS and revealing possibly interesting new physics from the tidal polarizability, two closely related questions need to be addressed timely. Firstly, what is actually the relationship between  the $\Lambda_{\rm{1.4}}$ and $R_{\rm{1.4}}$? In the literature, one often cites the proportionality $\Lambda \propto R^5 $ based on the definition $\Lambda = \frac{2}{3}k_2\cdot (R/M)^5 $. However, since the second Love number $k_2$ depends on $R$ for a given EOS thorough a very complicated differential equation coupled to the Tolman-Oppenheimer-Volkov (TOV) equation \cite{Hinderer08,Hinderer10}, the final  $\Lambda_{\rm{1.4}}-R_{\rm{1.4}}$ relation is not transparent {\it a priori} and the reported results in the literature are rather model dependent. For example, $\Lambda_{\rm{1.4}}\approx 7.76\times 10^{-4} \times R^{5.28}_{\rm{1.4}}$ was found using several energy density functionals within the Relativistic Mean Field (RMF) theory \cite{Fattoyev17}, while $\Lambda_{\rm{1.4}}\approx 9.11\times 10^{-5} \times R^{6.13}_{\rm{1.4}}$ was found from a systematic study using both the RMF and Skyrme Hartree-Fock (SHF) energy density functionals \cite{Malik}, and yet $\Lambda_{\rm{1.4}}\approx 2.88\times 10^{-6} \times R^{7.5}_{\rm{1.4}}$ was found based on a family of other EOSs constructed by interpolating between predictions of the state-of-the-art chiral effective field theory at low densities and the perturbative QCD at very high baryon densities using polytropes \cite{Ann}. Secondly, since the tidal polarizability will be precisely determined from the expected large number of GW observations of NS merger events in the near future (instead of only its upper limit or a large range currently available), it is interesting to know what aspects of the EOS of dense neutron-rich matter can be accurately determined by the $\Lambda_{\rm{1.4}}$. Given the close, albeit not exactly determined, relation between the $\Lambda_{\rm{1.4}}$ and $R_{\rm{1.4}}$, the second question is intemately related to the longstanding question of what information about the EOS can be extracted from the radii of NSs. While some interesting indications/speculations have been found/made, such as connections of the $\Lambda_{\rm{1.4}}$ with the slope parameter $L$ of nuclear symmetry energy $E_{\rm{sym}}(\rho )$ around the saturation density $\rho_0$, sizes of neutron-skins in heavy nuclei, and/or possible phase transitions in nuclear matter \cite{Fattoyev17,Xu09,Ang-quark,NBZ2018a}, much more work is still necessary to reveal the possible new physics from the $\Lambda_{\rm{1.4}}$ and/or $R_{\rm{1.4}}$. Indeed, some cautions have already been raised in the literature. For example, by comparing results of a systematic study on the  $\Lambda_{\rm{1.4}}$ and $R_{\rm{1.4}}$ using a quark-mean-field (QMF) model with predictions of several RMF, SHF, density-dependent relativistic Hartree-Fock (DDRHF) theories, as well as the microscopic Brueckner-Hartree-Fock (BHF) theory, the authors of \cite{Ang-quark} concluded that ``{\it there is no evidence for a simple relation between the symmetry energy slope L (hence the radius) and the tidal deformability. Consequently, claims regarding constraining NS radius with tidal deformability measurements should be considered with caution}''.

Besides the above considerations, it is well known that while the maximum mass of NSs is determined mostly by the EOS of isospin-symmetric nuclear matter, the radii of NSs with fixed masses are mainly determined by the density dependence of nuclear symmetry energy $E_{\rm{sym}}(\rho )$, see, e.g., \cite{LiSteiner}.  While it is customary to characterize the stiffness of nuclear symmetry energy with its slope parameter $L$ at the saturation density $\rho_0$, it is known that the radii of NSs are sensitive to the pressure of NS matter around twice the saturation density \cite{Lattimer01}. Moreover, the $L$ parameter alone is insufficient to characterize the entire density dependence of nuclear symmetry energy. While one often uses the single parameter $L$ to label predictions of different theories, it is obviously not unambiguous as some models may predict nuclear symmetry energies having the same $L$ but still quite different high-density behaviors, or vice versa \cite{Tesym}. Of course, predictions for the high-density behavior of nuclear symmetry energy based on nuclear many-body theories are usually correlated intrinsically with their predictions around the saturation density \cite{Roc11,Dutra1,Dutra2,ireview98,Bar05,Ste05,Li08,Tra12,Tsang12,Hor14,Bal16}, as often seen in examining the coefficients of Taylor expansions of nuclear energy density functionals. It is thus useful to know how different parts of the $E_{\rm{sym}}(\rho )$ may affect the $\Lambda_{\rm{1.4}}$ and $R_{\rm{1.4}}$ as well as their possible correlations.

\begin{figure*}
\begin{center}
\resizebox{1\textwidth}{!}{
  \includegraphics{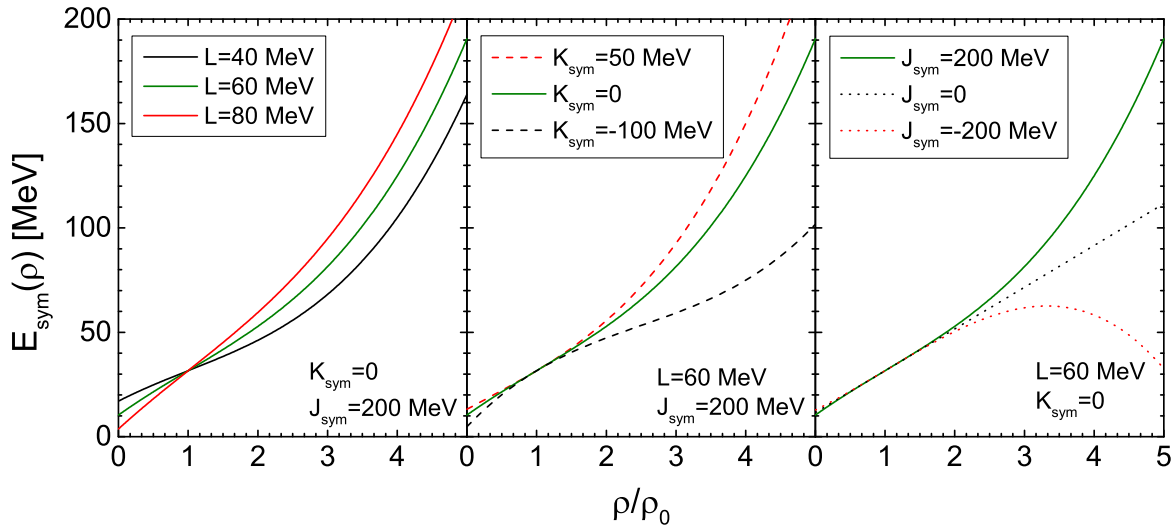}
}
\end{center}
\caption{(color online) Examples illustrating effects of the L (left), $K_{\rm{sym}}$ (middle) and $J_{\rm{sym}}$ (right), respectively, on the density dependence of nuclear symmetry energy $E_{\rm{sym}}(\rho )$. In each window,
only one of the parameters are varied.}
\label{Esym1}       

\end{figure*}

In this work, we first investigate effects of nuclear symmetry energy $E_{\rm{sym}}(\rho )$ on the radii and tidal polarizability of NSs and their correlations. We then study the NS inverse-structure problem of inferring constraints on the $E_{\rm{sym}}(\rho )$ parameters assuming the  $\Lambda_{\rm{1.4}}$ and $R_{\rm{1.4}}$ can be individually measured precisely.
For these purposes, we parameterize the symmetry energy as a function of density according to
\begin{equation}
E_{\rm{sym}}(\rho)=E_{\rm{sym}}(\rho_0)+L(\frac{\rho-\rho_0}{3\rho_0})+\frac{K_{\rm{sym}}}{2}(\frac{\rho-\rho_0}{3\rho_0})^2
+\frac{J_{\rm{sym}}}{6}(\frac{\rho-\rho_0}{3\rho_0})^3.\label{Esympara}
\end{equation}
Since the above parameterization naturally approaches asymptotically the Taylor expansion of $E_{\rm{sym}}(\rho )$ near the saturation density $\rho_0$, one can fix the parameters $E_{\rm sym}(\rho_0)$ and $L$ characterizing the $E_{\rm{sym}}(\rho )$ near $\rho_0$ at their known most probable values, while varying the high-density parameters $K_{\rm{sym}}$ and $J_{\rm{sym}}$ within their known uncertain ranges. Presently, the $K_{\rm{sym}}$ and $J_{\rm{sym}}$ are only known roughly to be around $-400 \leq K_{\rm{sym}} \leq 100$ MeV and $-200 \leq J_{\rm{sym}}\leq 800$ MeV \cite{Tews17,Zhang17}, while the most probable values of the $E_{\rm{sym}}(\rho_0)$ and $L$ have been relatively well constrained to be around $E_{\rm sym}(\rho_0)=31.7\pm 3.2$ MeV and $L=58.7\pm 28.1 $ MeV \cite{Oertel17,Li13}, respectively. Compared to existing studies based on nuclear many-body theories which often predict that different parts of the  $E_{\rm{sym}}(\rho )$ are generally correlated, namely, the $L$, $K_{\rm{sym}}$, and $J_{\rm{sym}}$ are not independent and their correlations are strongly model dependent, the parameterization of equation (\ref{Esympara}) enables us to examine effects of the $E_{\rm{sym}}(\rho )$ in different density regions on the radii and tidal polarizability. As shown in figure \ref{Esym1}, the $E_{\rm{sym}}(\rho )$ is varied above about $\rho_0$, 1.5$\rho_0$, and 2.5$\rho_0$ by varying individually the $L$ , $K_{\rm{sym}}$, and $J_{\rm{sym}}$ within their current uncertain ranges, respectively. We found that both the $\Lambda_{\rm{1.4}}$ and $R_{\rm{1.4}}$ vary with $L$ approximately linearly within its uncertain range with fixed $K_{\rm{sym}}$ and $J_{\rm{sym}}$. The $\Lambda_{\rm{1.4}}$ is more sensitive to the variations of the symmetry energy parameters especially the $K_{\rm{sym}}$ than the $R_{\rm{1.4}}$.  Assuming the $\Lambda_{\rm{1.4}}$ and $R_{\rm{1.4}}$ can be individually measured accurately, we show that they can both provide stringent and consistent constraints on the  $L$-$K_{\rm{sym}}$ relations.  Since infinite combinations of the larger (smaller) $L$ and smaller (larger) $K_{\rm{sym}}$ can lead to the same $\Lambda_{\rm{1.4}}$ and $R_{\rm{1.4}}$, additional observables are necessary to completely determine the density dependence of nuclear symmetry energy $E_{\rm{sym}}(\rho )$.

The paper is organized as follows. In the next section, we first summarize the major ingredients and justifications of our approach. Effects of the symmetry energy parameters on the $\Lambda_{\rm{1.4}}$, $R_{\rm{1.4}}$, and their correlations as well as results of solving the NS inverse-structure problem are presented in section \ref{Results}.  Finally, a summary of this work is given at the end.

\section{An explicitly isospin-dependent parametric EOS and the crust-core transition density in neutron star matter}\label{EOS}
For completeness and ease of our following discussions, we first summarize the main features of an explicitly isospin-dependent parametric EOS for NS matter. Here we only comment on features that are most important
for the study of NS radii. More details of constructing consistently the EOS of NS matter at $\beta$-equilibrium within this approach can be found in \cite{NBZ2018a,NBZ2018b}. The starting point is the empirical parabolic
law
\begin{equation}\label{EOS0}
E(\rho ,\delta )=E_0(\rho)+E_{\rm{sym}}(\rho )\cdot\delta ^{2} +\mathcal{O}(\delta^4)
\end{equation}
for the nucleon specific energy $E(\rho ,\delta )$ in nucleonic matter of density $\rho$ and isospin asymmetry $\delta\equiv (\rho_n-\rho_p)/\rho$,  see, e.g., \cite{Bom91}.
The nucleon specific energy $E_0(\rho)$ in symmetric nuclear matter (SNM) can be parameterized as
\begin{equation}\label{E0para}
E_{0}(\rho)=E_0(\rho_0)+\frac{K_0}{2}(\frac{\rho-\rho_0}{3\rho_0})^2+\frac{J_0}{6}(\frac{\rho-\rho_0}{3\rho_0})^3.
\end{equation}
While the incompressibility $K_0$ of SNM has been constrained to $K_0=230 \pm 20$ MeV \cite{Shlomo06,Piekarewicz10}, the high-density parameter $J_0$ is roughly known to be around $-800 \leq J_{0}\leq 400$ MeV \cite{Tews17,Zhang17}. Interestingly, electromagnetic constraints on the remnant imposed by the kilonova observations together with the gravitational wave information \cite{Margalit17} have recently enabled a constraint of M$_{\rm{max}}\leq 2.17$ M$_\odot$ with 90\% confidence on the maximum mass of NSs. Moreover, using general considerations and universal relations without using directly any simulation, the GW170817 was also found to point to the same maximum mass of 2.17 M$_\odot$ \cite{Rezzolla18,Rez2}. This new constraint on the NS maximum mass and the previously observed NS maximum mass of 2.01 M$_\odot$
have very recently enabled us to limit the $J_0$ in the range of about $-200\pm 25$ MeV \cite{NBZ2018b}. Since we are focusing on the radius and tidal polarizability of canonical NSs,
we fix the value of $J_0$ at $-180$ MeV and focus on effects of the symmetry energy. Our conclusions regarding the $\Lambda_{\rm{1.4}}$ and $R_{\rm{1.4}}$ (which have little dependence on the EOS of SNM) are qualitatively independent of this choice.

It is well known that the crust affects significantly the radius while it contributes little to the total mass of a NS. Thus, the crust-core transition density/pressure is an important quantity in determining the radii of NSs, especially those of low mass NSs. As discussed in detail in \cite{NBZ2018a}, we connect self-consistently and smoothly the core EOS with the NV EOS \cite{Negele73} for the inner crust and the BPS EOS  \cite{Baym71} for the outer crust. As illustrated in figure \ref{Esym1}, the variations of $L$ and $K_{\rm{sym}}$ affect significantly the sub-saturation behavior of $E_{\rm{sym}}(\rho)$, and thus the crust-core transition density. We also notice here that the free variations of  $L$ and $K_{\rm{sym}}$ may lead to non-zero $E_{\rm{sym}}(\rho)$ at the $\rho=0$ limit. This purely mathematical limit has no physical effect on NSs in our study as the EOS in the very low density region is replaced by the BPS EOS for the outer crust. It is also worth noting that for clustered matter, the symmetry energy is known to have a non-zero asymptotic value at $\rho=0$ both experimentally \cite{Joe} and theoretically \cite{Typel}. However, we are not aware of any physical reason for the symmetry energy of uniform matter to be non-zero at the $\rho=0$ limit. While there are precise predictions for the EOS of pure neutron matter at very low densities where symmetric nuclear matter is unstable against cluster formation and pairing, the symmetry energy for uniform matter at densities less than the NS crust-core transition density thus has no practical applications in studying neutron stars. 
We also notice that thermodynamical stability conditions require that the pressure stays positive and always increases with increasing density. Thus, in varying the parameters we require the crust-core transition density to stay positive. This limits the allowed parameter space.

\begin{figure*}
\resizebox{0.45\textwidth}{6cm}{
  \includegraphics{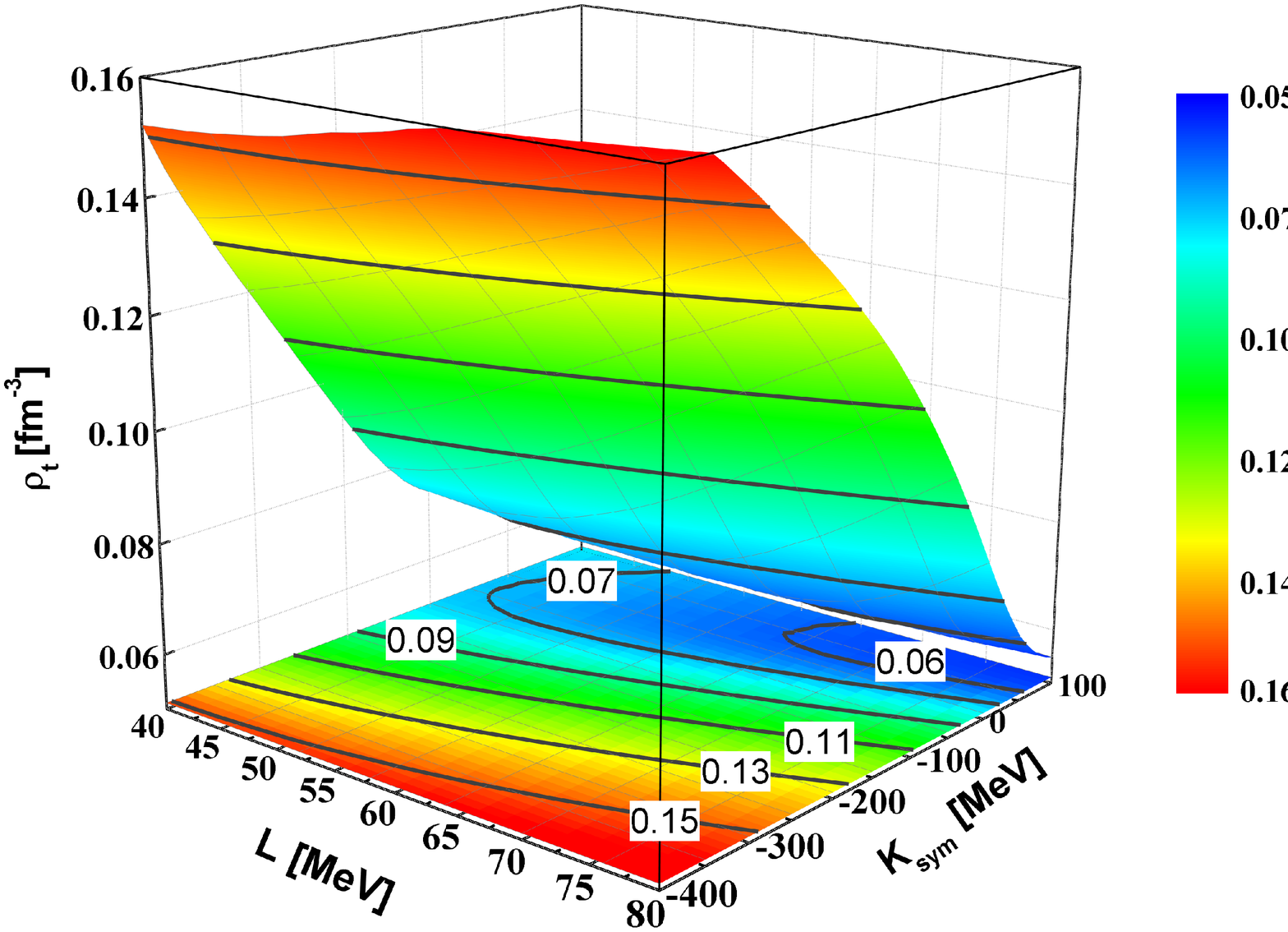}
  }
  \hspace{0.1cm}
  \resizebox{0.45\textwidth}{6cm}{
 \includegraphics{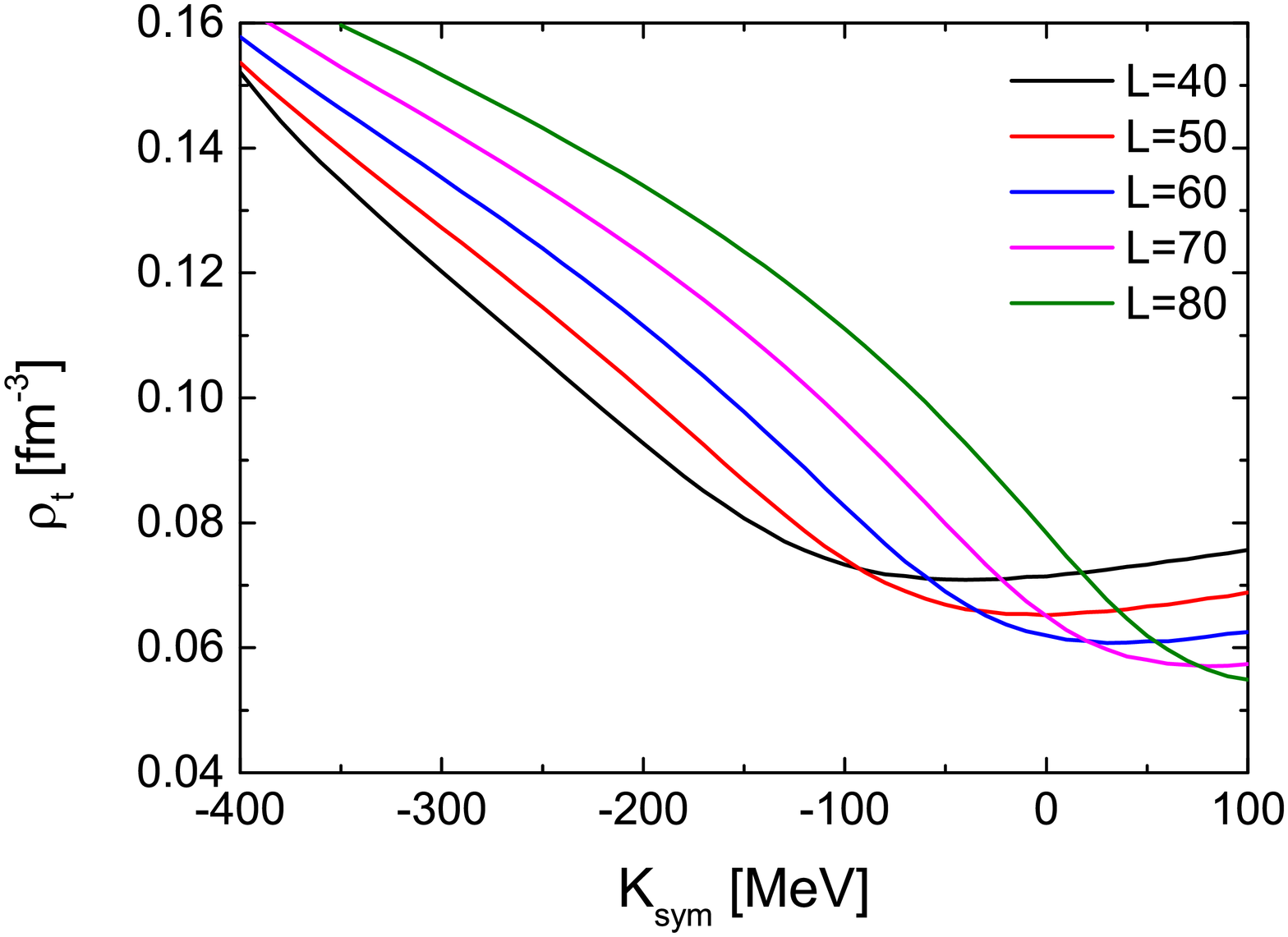}
}
  \caption{(color online) Effects of nuclear symmetry energy on the crust-core transition density in NSs. The left window is the transition density as a function of $L$ and $K_{\rm{sym}}$. The projection of the
  transition density is also shown as contours in the $L$-$K_{\rm{sym}}$ plane. The right window (taken from \cite{NBZ2018a}) is the transition density as a function of $K_{\rm{sym}}$ with fixed $L$ values.
  }
\label{Tdensity}       
\end{figure*}

The core-crust transition density was found by examining the incompressibility of NS matter
\begin{equation}\label{tPA}
K_\mu=\rho^2\frac{d^2E_0}{d\rho^2}+2\rho\frac{dE_0}{d\rho}+\delta^2\left[\rho^2\frac{d^2E_{\rm{sym}}}{d\rho^2}+2\rho\frac{dE_{\rm{sym}}}{d\rho}-2E^{-1}_{\rm{sym}}(\rho\frac{dE_{\rm{sym}}}{d\rho})^2\right]
\end{equation}
for any given set of EOS parameters \cite{Kubis04,Kubis07,Lattimer07}. Once the $K_\mu$ becomes negative, the thermodynamical instability will grow by forming clusters, indicating a transition from the uniform core to the clustered crust. It is seen that the $K_\mu$ involves the first-order and second-order derivatives of the symmetry energy, i.e., quantities related to the $L$ and $K_{\rm{sym}}$. In fact, the last two terms in the bracket of the above expression are largely canceled out, leaving the $K_{\rm{sym}}$ dominates. Shown in the left window of figure \ref{Tdensity} is the transition density as a function of $L$ and $K_{\rm{sym}}$. The projection of the transition density is also shown as contours in the $L$-$K_{\rm{sym}}$ plane. To more clearly reveal the role of $K_{\rm{sym}}$ with respect to that of $L$, shown in the right window is the transition density as a function of $K_{\rm{sym}}$ for a given $L$. It is clearly seen that the $K_{\rm{sym}}$ plays a more significant role in determining the crust-core transition. In some sense, this is natural as the curvature of the symmetry energy is a major part of the incompressibility of NS matter. This also indicates that it is insufficient to differentiate predicted $E_{\rm{sym}}(\rho)$ with only the $L$ parameter and study only effects of $L$ on the radii of NSs. In fact, not only through the pressure at supra-saturation densities around $2\rho_0$, the $K_{\rm{sym}}$ affects the radii also through its influences on the crust-core transition density. Indeed, as we shall demonstrate next, the $K_{\rm{sym}}$ affects the radii appreciably for a given $L$.

\section{Effects of nuclear symmetry energy on the radii and tidal polarizabilities of neutron stars as well as their correlations}\label{Results}
\begin{figure*}
\begin{center}
\resizebox{1\textwidth}{!}{
  \includegraphics{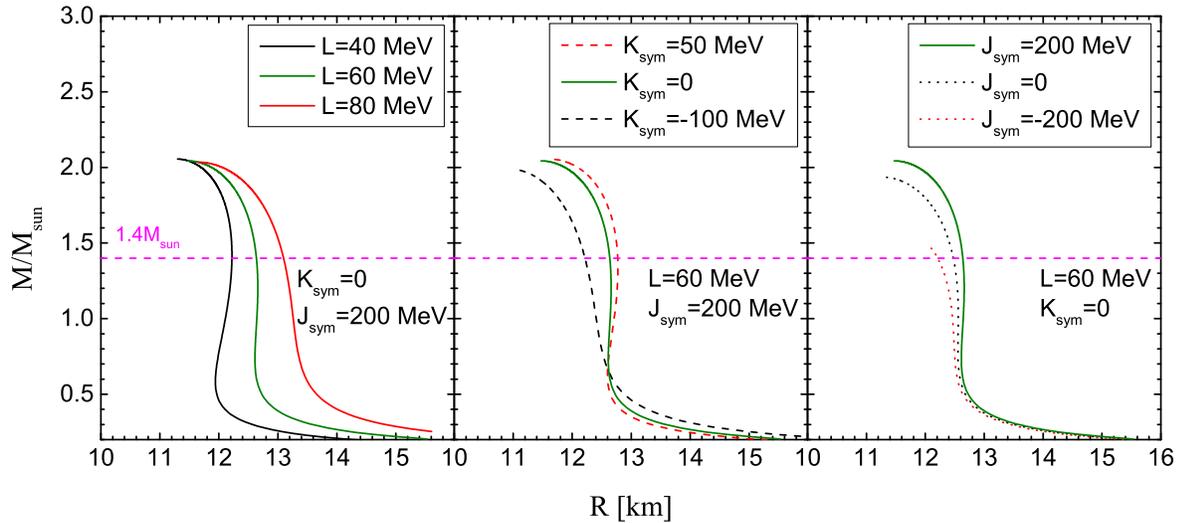}
  }
 \caption{(color online) Examples illustrating effects of the $L$, $K_{\rm{sym}}$ and $J_{\rm{sym}}$, respectively, on the mass-radius correlation of NSs. }
\label{MR}       
\end{center}
\end{figure*}

Within the uncertainty ranges of the three parameters $L$, $K_{\rm{sym}}$, and $J_{\rm{sym}}$, in this section we investigate effects of the symmetry energy
on the radii and tidal polarizabilities of NSs. Typical examples of the $E_{\rm{sym}}(\rho)$ from varying individually the three parameters are shown in figure \ref{Esym1}.
It is known that variations of parameters characterizing the EOS of SNM, such as the $K_0$ and $J_0$, have negligible effect on the $\Lambda_{\rm{1.4}}$ and $R_{\rm{1.4}}$ \cite{NBZ2018b}.
Here we set $K_0=230 $ MeV and $J_0=-180$ MeV as we discussed earlier. Shown in figure \ref{MR} are the NS mass-radius relations by varying the $L$ (left), $K_{\rm{sym}}$ (middle), and $J_{\rm{sym}}$ (right), respectively. In the left window, we vary $L$ around its most probable value of $L=60$ MeV. With the $K_{\rm{sym}}$ set to zero, the $J_{\rm{sym}}$ is set to 200 MeV so that the resulting EOS can support
NSs as heavy as 2.01 M$_\odot$. In the middle and right windows, individual effects of varying the $K_{\rm{sym}}$ and $J_{\rm{sym}}$ are studied while the $L$ is fixed at its most probable value of $L=60$ MeV.
It is seen from the left window that the $R_{\rm{1.4}}$ changes from 12.2 km to 13.1 km, representing a 7\% variation when the value of $L$ is changed by 50\% (with respect to the highest $L$ considered) from 40 to 80 MeV. With the $L$ fixed as 60 MeV, it is seen from the middle and right widows that variations of $K_{\rm{sym}}$ by 150\% and $J_{\rm{sym}}$ by 200\% lead, respectively, to about a 5\% and 3\% change in the $R_{\rm{1.4}}$. Thus the $R_{\rm{1.4}}$ is most strongly affected by the $L$ value (28\%) while effects of the $K_{\rm{sym}}$ (7\%) and $J_{\rm{sym}}$ (3\%) are appreciable if their effects are measured with respect to the same relative change of all three variables.  It is also interesting to note in the right window that, as a high-density parameter, the $J_{\rm{sym}}$ has a more appreciable effect on the NS maximum mass than on the radius $R_{\rm{1.4}}$.

\begin{figure*}
\begin{center}
\resizebox{1\textwidth}{!}{
  \includegraphics{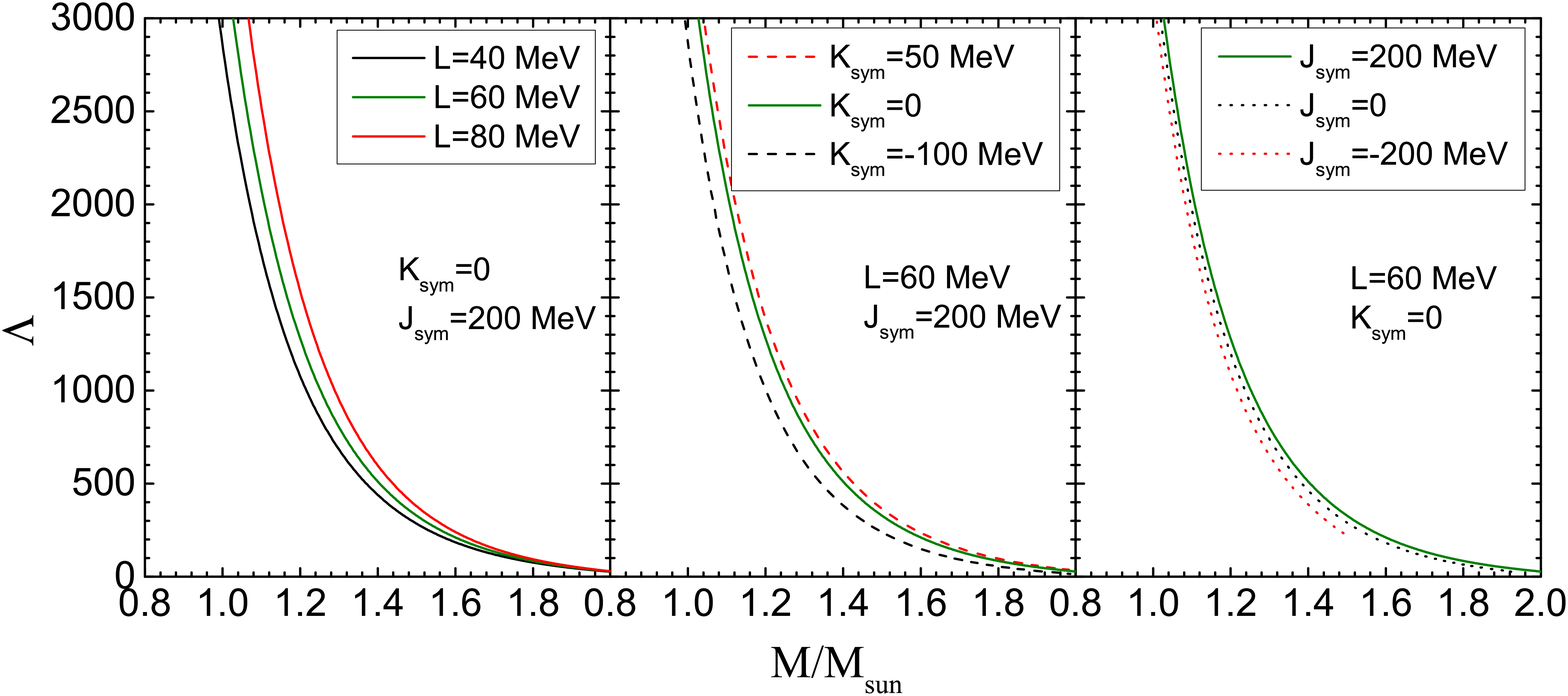}
  }
 \caption{(color online) Examples illustrating effects of the $L$, $K_{\rm{sym}}$ and $J_{\rm{sym}}$, respectively, on the mass-tidal polarizability correlation of NSs. }
\label{LM}       
\end{center}
\end{figure*}

\begin{figure*}
\begin{center}
\resizebox{1\textwidth}{!}{
  \includegraphics{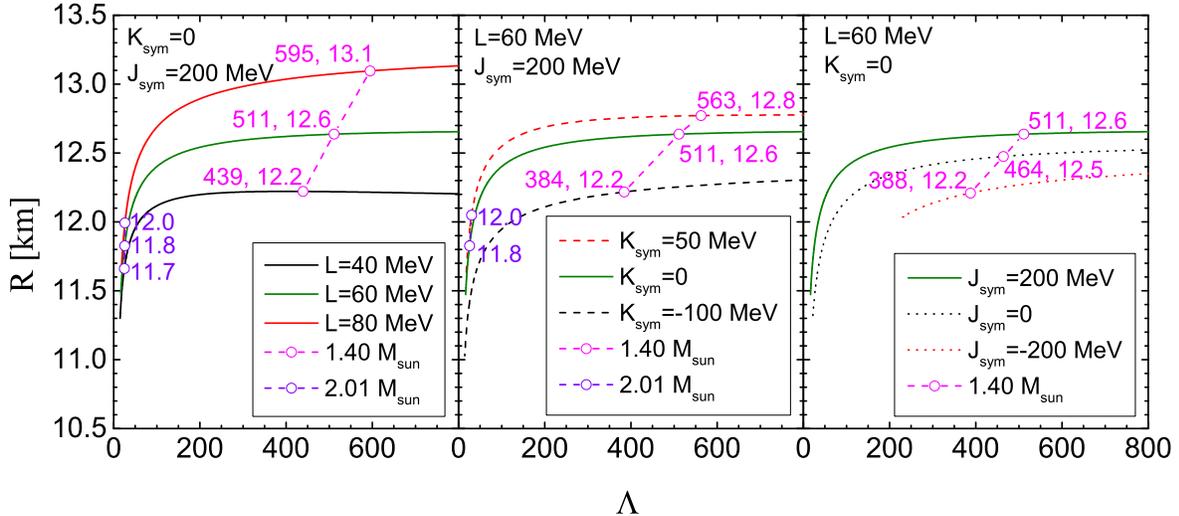}
  }
  \caption{(color online) Examples illustrating effects of the $L$, $K_{\rm{sym}}$ and $J_{\rm{sym}}$, respectively, on the radius-tidal polarizability correlation of NSs. }
\label{RL}       
\end{center}
\end{figure*}
The corresponding variations of the tidal polarizability $\Lambda$ with $L$, $K_{\rm{sym}}$, and $J_{\rm{sym}}$ as functions of mass are shown in figure \ref{LM}. At 1.4 M$_\odot$, the un-normalized relative change of $\Lambda_{1.4}$ is about 36\%, 47\% and 32\% corresponding to the variations of L (50\%), $K_{\rm{sym}}$ (150\%) and $J_{\rm{sym}}$ (200\%), respectively. They are indeed much larger than the variations of $R_{\rm{1.4}}$. These results thus verify the expectation that the tidal polarizability is more sensitive to the variation of $E_{\rm{sym}}(\rho)$ than the radius. The correlations between the radius and tidal polarizability with the same EOSs are shown in figure \ref{RL}. Along each curve from the left to right, the NS mass decreases while the radius first increases then stays approximately a constant around $M=1.4$ M$_{\odot}$. For NSs with a fixed mass, such as M=2.01 M$_{\odot}$ (violet) and M=1.4 M$_{\odot}$ (magenta), in the left window, the radius increases approximately linearly with $\Lambda$. For example, as $L$ varies from 40 to 60 to 80 MeV, the $R_{\rm{1.4}}$ changes from 12.2 to 12.6 to 13.1 km while the $\Lambda_{1.4}$ changes from 439 to 511 to 595.  Albeit at slightly different rates, the variations of $K_{\rm{sym}}$ and $J_{\rm{sym}}$ also lead to approximately linear variations of $R_{\rm{1.4}}$ and $\Lambda_{1.4}$.
We notice that the highly non-linear and significantly different $\Lambda_{1.4}-R_{\rm{1.4}}$ correlations found in \cite{Ann,Fattoyev17,Malik} were obtained in much larger ranges of $R_{\rm{1.4}}$ ($\Lambda_{1.4}$) compared to what we found here using the relatively narrowly constrained L values. More quantitatively, they calculated the $R_{\rm{1.4}}$ ($\Lambda_{1.4}$) in approximately the ranges of 7.0-14.5 km (0-1600) \cite{Ann},
12.5-14.7 km (500-1300) \cite{Fattoyev17} and 11.5-14.7 km (300-1200) \cite{Malik}, respectively. Of course, in smaller regions of $R_{\rm{1.4}}$ along their curves, the $R_{\rm{1.4}}-\Lambda_{1.4}$ correlations look more linear. Nevertheless, the exact $R_{\rm{1.4}}-\Lambda_{1.4}$ correlation remains model dependent and certainly deserves further studies.

\section{Inferring the density dependence of nuclear symmetry energy from precisely measured radii and tidal polarizabilities of canonical neutron stars}\label{Results2}
\begin{figure*}
\begin{center}
\resizebox{0.49\textwidth}{6cm}{
  \includegraphics{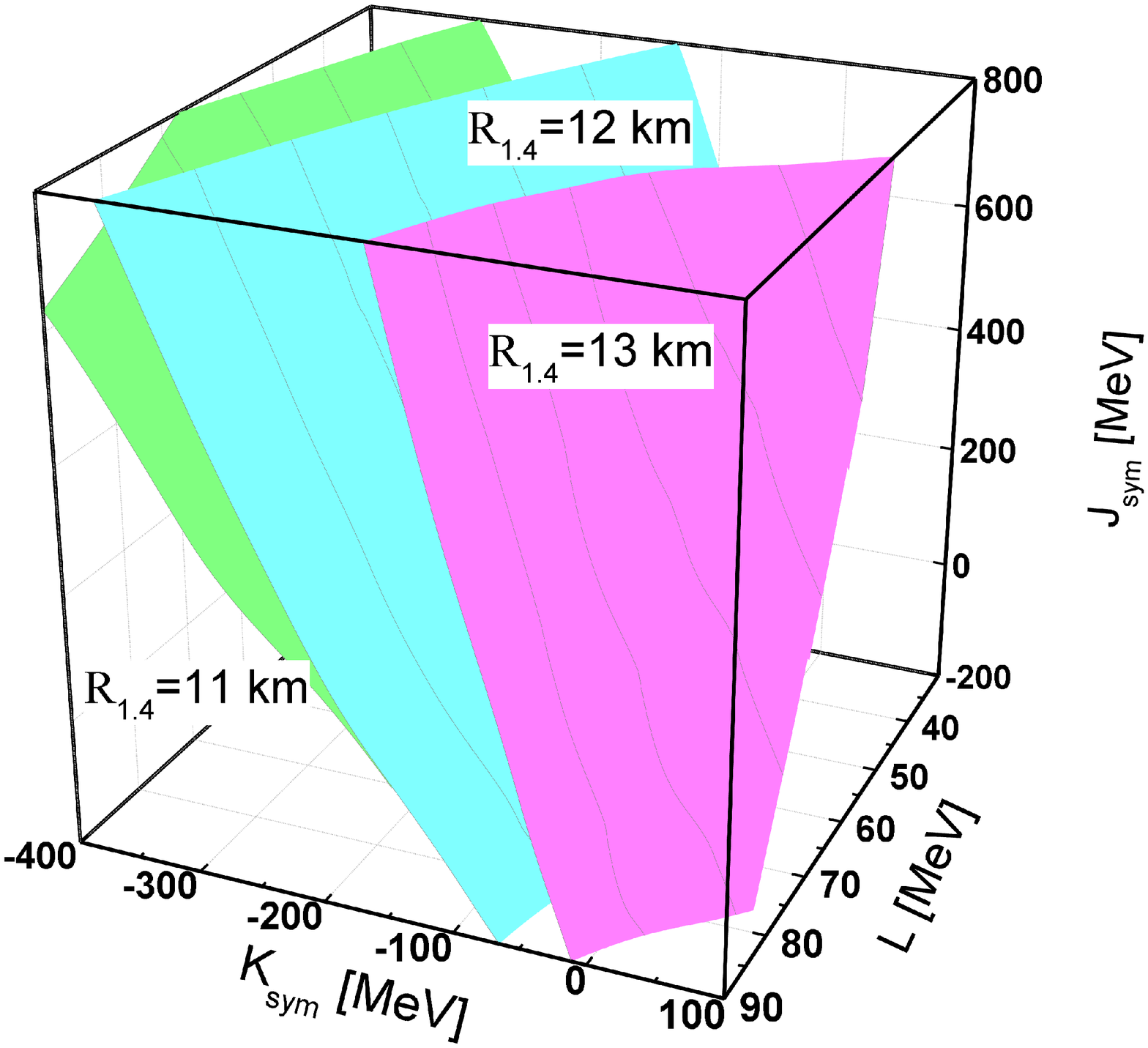}
  }
  \resizebox{0.49\textwidth}{6cm}{
 \includegraphics{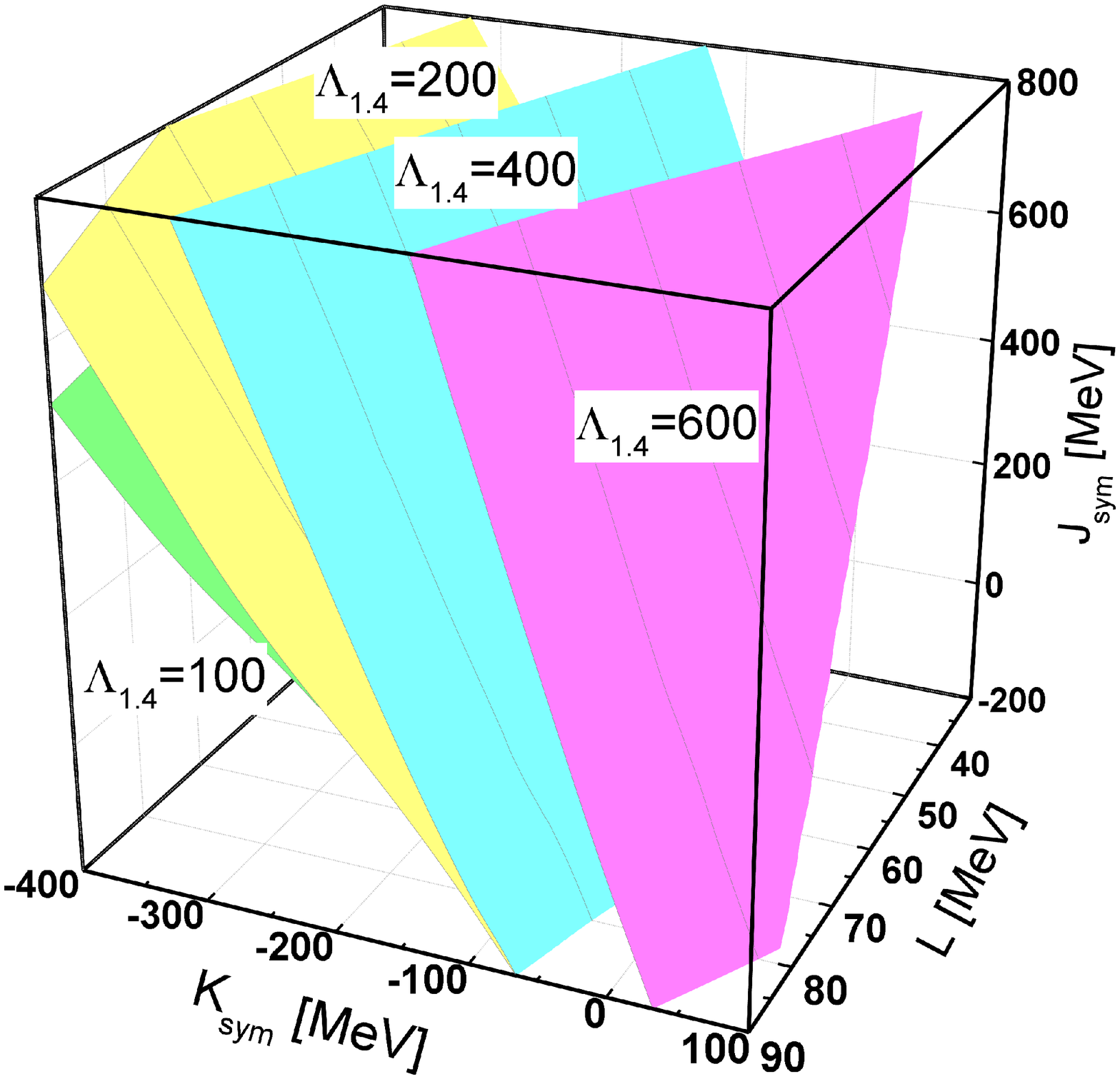}
}

  \caption{(color online) Surfaces of constant radius (left) and tidal polarizability (right) in the symmetry energy parameter space of $L$, $K_{\rm{sym}}$, and $J_{\rm{sym}}$, respectively. }
\label{RL-3D}       
\end{center}
\end{figure*}
What information about the $E_{\rm{sym}}(\rho)$ can be extracted assuming the $R_{\rm{1.4}}$ and $\Lambda_{1.4}$ can be observationally measured accurately? This is an inverse-structure problem and the answer may have interesting implications. We have shown in an earlier work \cite{NBZ2018b} that such kind of inverse-structure problems can be solved numerically using our parameterized EOS in integrating the TOV equation. Shown in the two blocks of figure \ref{RL-3D} are the constant surfaces of $R_{\rm{1.4}}$ and $\Lambda_{1.4}$ in the three-dimensional (3D) parameter space of nuclear symmetry energy. These surfaces represent necessary combinations of the $L$, $K_{\rm{sym}}$, and $J_{\rm{sym}}$ parameters to produce the specified values of $R_{\rm{1.4}}$ or $\Lambda_{1.4}$.  The whole 3D space shown is currently allowed by the known uncertainties of the three parameters. As one expects, only one observable, either the $R_{\rm{1.4}}$ or $\Lambda_{1.4}$, is insufficient to completely determine the three parameters but provides a strong constraint on them. As examples, three surfaces with $R_{\rm{1.4}}=11, 12$, and 13 km and four surfaces with $\Lambda_{1.4}=100, 200, 400$, and 600, respectively, are shown in the two blocks.  Because the central density reached in canonical NSs is not very high, the high-density parameter $J_{\rm{sym}}$ plays little role in determining the radii of these NSs. Namely, the $J_{\rm{sym}}$ can be essentially any value in its uncertain range without affecting the value of $R_{\rm{1.4}}$. On the other hand, while the $L$ dominates, the $K_{\rm{sym}}$ has an appreciable role in determining the radius as we have already seen in the predicted M-R relation given a set of the $E_{\rm{sym}}(\rho)$ parameters in figure \ref{MR}. In the inverse-structure problem here, for $R_{\rm{1.4}}$=12 km as an example, it can be obtained with a large $L$ but small $K_{\rm{sym}}$ on one end or a small $L$ but larger $K_{\rm{sym}}$ on the other end. Thus, a precise measurement of $R_{\rm{1.4}}$ alone can only constrain a combination of the $L$ and $K_{\rm{sym}}$. As shown in the right block, an independent measurement of $\Lambda_{1.4}$ would provide similar information about the combination of $L$ and $K_{\rm{sym}}$ parameters. Since the $R_{\rm{1.4}}$ and $\Lambda_{1.4}$ are intrinsically correlated, independent measurements of them would provide complementary information and enable useful cross checks on the inferred symmetry energy $E_{\rm{sym}}(\rho)$. 

To this end, it is also useful to stress that in solving the NS inverse-structure problem of inferring the $E_{\rm{sym}}(\rho)$ parameters from precise measurements of $R_{\rm{1.4}}$ and $\Lambda_{1.4}$, we are not using much prior information about the L except its range. Essentially, we assumed that all L values are equally possible, equivalent to using a flat prior probability density distribution for the L in the Bayesian language. In fact, extensive studies over the last two decades mostly using data from terrestrial nuclear experiments have indicated that the most probable value of L is about 60 MeV, albeit still has a large width/uncertainty \cite{Oertel17,Li13}.  If one adopts the most probable value of L, the correlation between $L$ and $K_{\rm{sym}}$ parameters will be broken. Such an approach was used in our previous work to set an upper limit on the high-density behavior of $E_{\rm{sym}}(\rho)$ \cite{NBZ2018b}.

We notice that the constant surfaces of $\Lambda_{1.4}$ become more inclined in the negative $K_{\rm{sym}}$ direction towards $\Lambda_{1.4}$=100, indicating the need for a softer $E_{\rm{sym}}(\rho)$ at high densities with smaller values of both $K_{\rm{sym}}$ and $J_{\rm{sym}}$. This is completely understandable from inspecting again the results shown in figure \ref{RL}. For a fixed mass of 1.4 M$_\odot$, a smaller value of $\Lambda_{1.4}$ requires a smaller radius $R_{\rm{1.4}}$. The results in both figures \ref{RL} and the left block of figure \ref{RL-3D} indicate that smaller radii can be obtained by using smaller values of $K_{\rm{sym}}$ and $J_{\rm{sym}}$ without changing the value of $L$. Again, it indicates that the value of $L$ alone can not uniquely determine the radii of NSs. Information about the high-density behavior of $E_{\rm{sym}}(\rho)$ is also important.
\begin{figure*}
\begin{center}
\resizebox{0.45\textwidth}{!}{
  \includegraphics{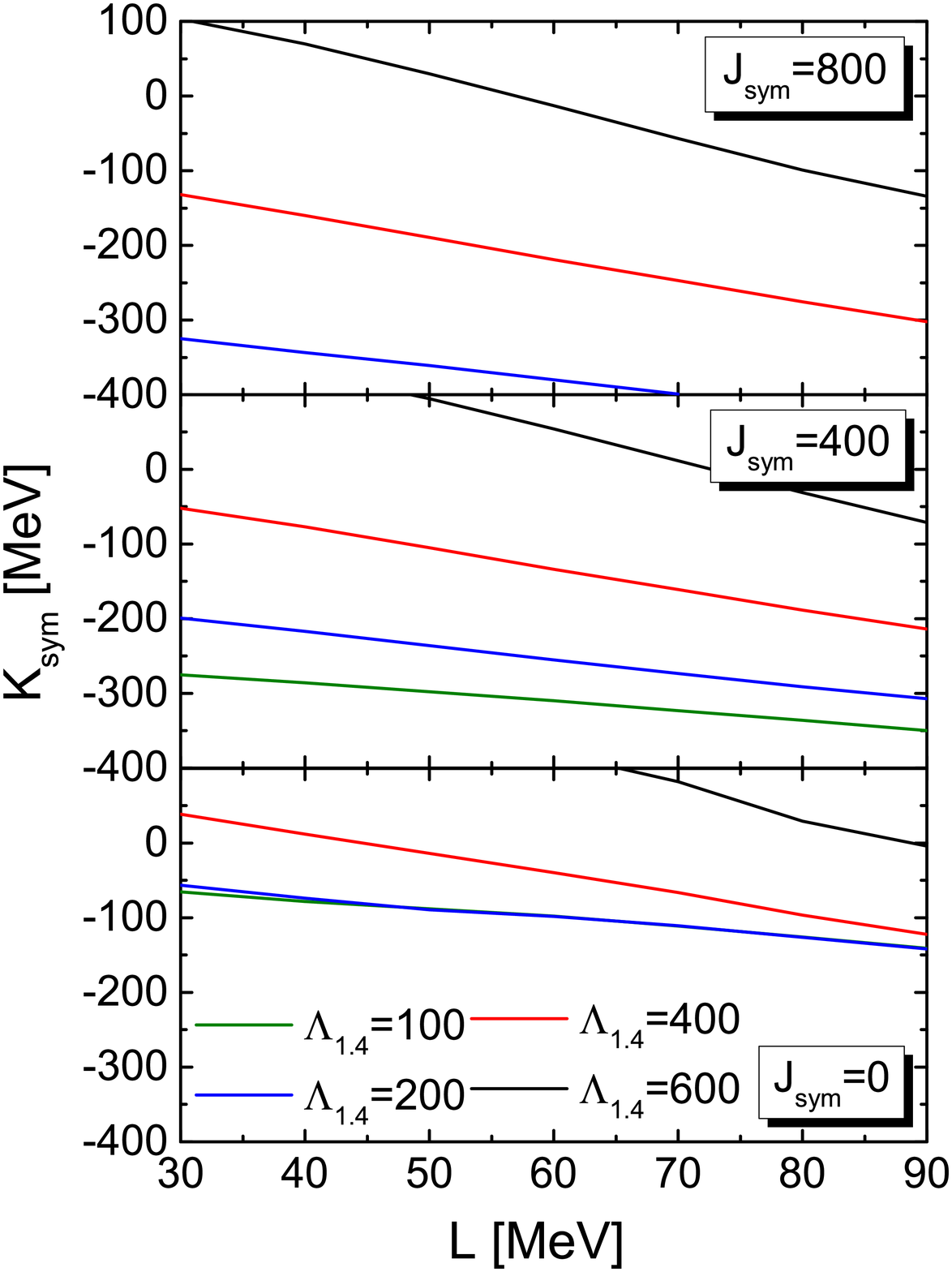}
  }
  \resizebox{0.45\textwidth}{!}{
 \includegraphics{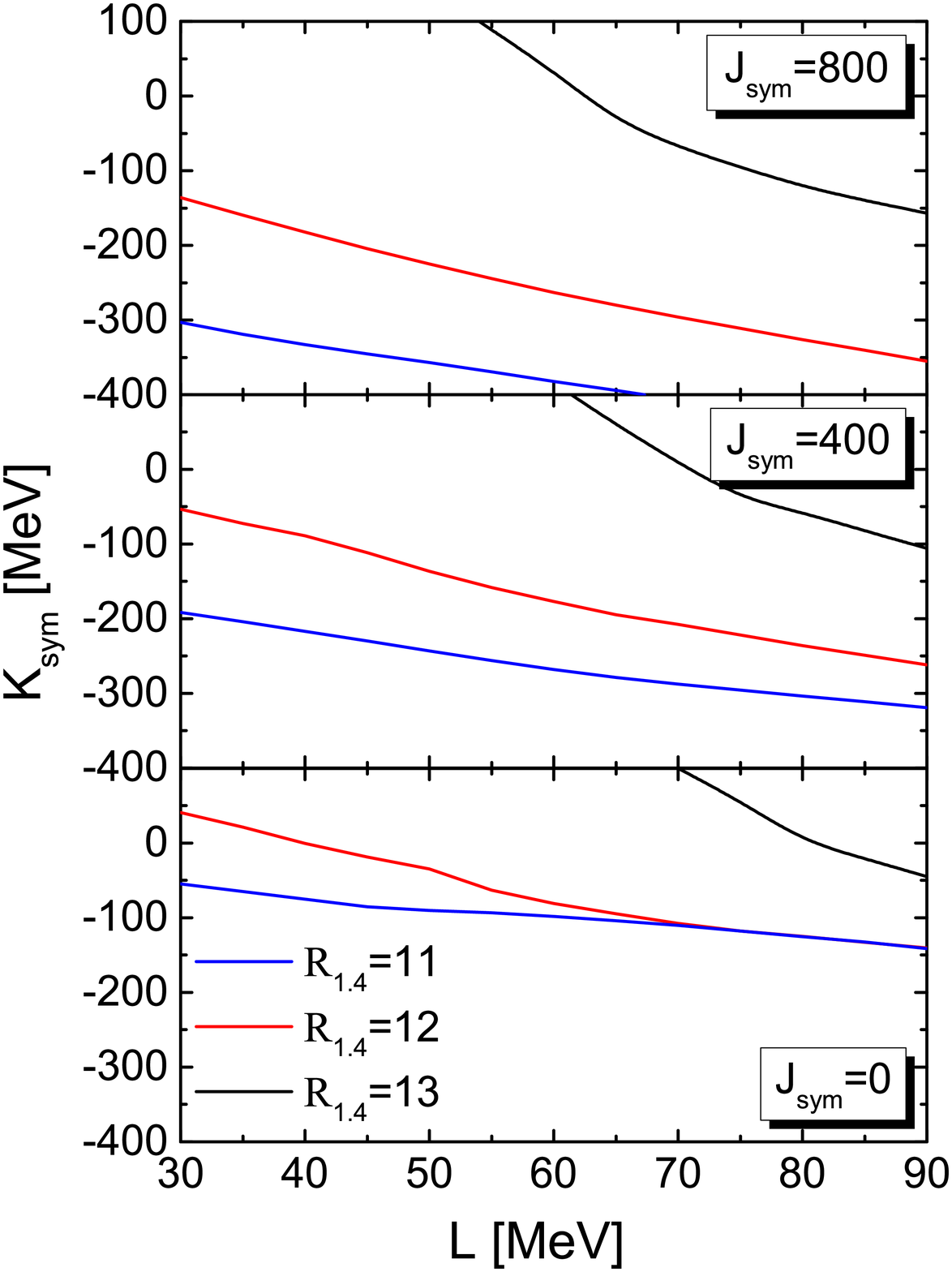}
}
\caption{(color online) Contours of tidal polarizability (left) and radius (right) in the $L$-$K_{\rm{sym}}$ plane with constant values of $J_{\rm{sym}}$. }
\label{RL-2D}       
\end{center}
\end{figure*}

To be more quantitative about the constraints on the $L$-$K_{\rm{sym}}$ correlations set by the expected precise measurements of $R_{\rm{1.4}}$ and $\Lambda_{1.4}$, shown in figure \ref{RL-2D} are contours of
$\Lambda_{1.4}$ and $R_{\rm{1.4}}$ in the $L$-$K_{\rm{sym}}$ plane for a given $J_{\rm{sym}}$. Again, the latter has little effect except for very small values of $\Lambda_{1.4}$.
Overall, for any given value of $R_{\rm{1.4}}$ or $\Lambda_{1.4}$, the required value of $K_{\rm{sym}}$ decreases with increasing $L$. As we discussed earlier, a stiffer $E_{\rm{sym}}(\rho)$ at saturation density with a larger $L$ but softer $E_{\rm{sym}}(\rho)$ at higher densities with a smaller $K_{\rm{sym}}$ leads to the same $R_{\rm{1.4}}$ and $\Lambda_{1.4}$ as a softer $E_{\rm{sym}}(\rho)$ at saturation density with a smaller $L$ but stiffer $E_{\rm{sym}}(\rho)$ at higher densities with a larger $K_{\rm{sym}}$. Thus, the often used practice of labeling predicted $E_{\rm{sym}}(\rho)$ with L alone and comparing the corresponding $R_{\rm{1.4}}$ and $\Lambda_{1.4}$ is insufficient. Indeed, ``{\it there is no evidence for a simple relation between the symmetry energy slope $L$ (hence the radius) and the tidal deformability \cite{Ang-quark}."} Moreover, by definition, the $K_{\rm{sym}}$ reflects the change of $L$. The possible observation of a stiffer $E_{\rm{sym}}(\rho)$ at sub-saturation densities from terrestrial experiments, such as the sizes of neutron skins of heavy nuclei, and a softer $E_{\rm{sym}}(\rho)$ at supra-saturation densities from NS radius or tidal polarizability measurements is not necessarily an indication of any phase transition. They may simply need a negative value of $K_{\rm{sym}}$ at saturation density or different combinations of the $L$ and $K_{\rm{sym}}$ parameters at sub-saturation and supra-saturation densities of purely nucleonic matter.

\section{Summary}
In summary, the reported NS tidal polarizability from the first multi-messenger observation of the binary NS merger event GW170817 has stimulated much interest in studying the EOS of dense neutron-rich matter. In this still earlier phase of the study, there are already many interesting findings. Our work here shed new light on a few less clearly studied questions regarding what we can learn from the NS radius and/or tidal polarizability about the density dependence of nuclear symmetry energy $E_{\rm{sym}}(\rho )$. The latter is presently the most uncertain term in the EOS of dense neutron-rich nucleonic matter. Using an explicitly isospin-dependent EOS with the $E_{\rm{sym}}(\rho )$ parameterized using the three parameters $L$, $K_{\rm{sym}}$ and $J_{\rm{sym}}$, we first investigated effects of these parameters on the NS radius and tidal polarizability as well as their correlations. It is found that while both the $R_{\rm{1.4}}$ and $\Lambda_{1.4}$ depend strongly on $L$, the high-density behavior of $E_{\rm{sym}}(\rho )$ characterized by the $K_{\rm{sym}}$ and $J_{\rm{sym}}$ parameters plays appreciable roles. The $R_{\rm{1.4}}$ and $\Lambda_{1.4}$ are approximately linearly correlated in the currently known uncertainty ranges of the $E_{\rm{sym}}(\rho )$ parameters.  Moreover, the tidal polarizability is found to be more sensitive to the variation of $E_{\rm{sym}}(\rho)$ than the radius. We also studied the inverse-structure problem of inferring constraints on the $E_{\rm{sym}}(\rho )$ parameters assuming the  $\Lambda_{\rm{1.4}}$ and $R_{\rm{1.4}}$ can be individually measured precisely.
We found that the individual measurements of $\Lambda_{\rm{1.4}}$ and $R_{\rm{1.4}}$ can not completely determine the $E_{\rm{sym}}(\rho )$ but can limit combinations of its parameters. In particular, stringent constraints on the $L$-$K_{\rm{sym}}$ correlation can be obtained. However, there is not a simple relation between the $\Lambda_{\rm{1.4}}$/$R_{\rm{1.4}}$ and $L$ alone. Generally speaking, many combinations of the larger (smaller) $L$ and smaller (larger) $K_{\rm{sym}}$ within their existing constraints mostly from analyzing terrestrial nuclear experiments can lead to the same $\Lambda_{\rm{1.4}}$ and $R_{\rm{1.4}}$. Thus, additional observables including those from terrestrial nuclear experiments are necessary to break this degeneracy in order to completely determine the density dependence of nuclear symmetry energy $E_{\rm{sym}}(\rho )$.

\section*{Acknowledgement}
We would like to thank Lie-Wen Chen, Plamen G. Krastev, Bin Qi, De-Hua Wen, and Jun Xu for helpful discussions. NBZ was supported in part by the China Scholarship Council. BAL acknowledges the U.S. Department of Energy, Office of Science, under Award Number DE-SC0013702, the CUSTIPEN (China-U.S. Theory Institute for Physics with Exotic Nuclei) under the US Department of Energy Grant No. DE-SC0009971 and the National Natural Science Foundation of China under Grant No. 11320101004.

%

\end{document}